\documentclass[aip,graphicx]{revtex4-1}
\usepackage[cp1251]{inputenc}
\usepackage[T2A]{fontenc}
\usepackage[english]{babel}
\usepackage{amssymb,latexsym,amsmath,amscd}
\usepackage{graphicx,color,framed}
\begin{document}
\selectlanguage{english}
\newcommand{\angstrom}{\text{\normalfont\AA}}

\title{On the theory of electric double layer with explicit account of a polarizable co-solvent}

\author{\firstname{Yu.~A.} \surname{Budkov}}
\email[]{urabudkov@rambler.ru}
\affiliation{G.A. Krestov Institute of Solution Chemistry of the Russian Academy of Sciences, Laboratory of NMR spectroscopy and numerical investigations of liquids, Ivanovo, Russia}
\affiliation{National Research University Higher School of Economics, Department of Applied Mathematics, Moscow, Russia}

\author{\firstname{ A.~L.} \surname{Kolesnikov}}
\affiliation{Institut f\"{u}r Nichtklassische Chemie e.V., Universit\"{a}t Leipzig, Leipzig, Germany}

\author{\firstname{ M.~G.} \surname{Kiselev}}
\affiliation{G.A. Krestov Institute of Solution Chemistry of the Russian Academy of Sciences, Laboratory of NMR spectroscopy and numerical investigations of liquids, Ivanovo, Russia}
\begin{abstract}
We present a continuation of our theoretical research into the influence of co-solvent polarizability on a differential capacitance of the electric double layer. We formulate a modified Poisson-Boltzmann theory, using the formalism of density functional approach on the level of local density approximation taking into account the electrostatic interactions of ions and co-solvent molecules as well as their excluded volume. We derive the modified Poisson-Boltzmann equation, considering the three-component symmetric lattice gas model as a reference system and minimizing the grand thermodynamic potential with respect to the electrostatic potential. We apply present modified Poisson-Boltzmann equation to the electric double layer theory, showing that accounting for the excluded volume of co-solvent molecules and ions slightly changes the main result of our previous simplified theory. Namely, in the case of small co-solvent polarizability with its increase under the enough small surface potentials of electrode the differential capacitance undergoes the significant growth. Oppositely, when the surface potential exceeds some threshold value (which is slightly smaller than the saturation potential), the increase in the co-solvent polarizability results in a differential capacitance decrease. However, when the co-solvent polarizability exceeds some threshold value, its increase generates a considerable enhancement of the differential capacitance in a wide range of surface potentials. We demonstrate that two qualitatively different behaviors of the differential capacitance are related to the depletion and adsorption of co-solvent molecules at the charged electrode. We show that an additive of the strongly polarizable co-solvent to an electrolyte solution can shift significantly the saturation potential in two qualitatively different manners. Namely, a small additive of strongly polarizable co-solvent results in a shift of saturation potential to higher surface potentials. On the contrary, a sufficiently large additive of co-solvent shifts the saturation potential to lower surface potentials. We obtain that an increase in the co-solvent polarizability makes the electrostatic potential profile longer-ranged. However, increase in the co-solvent concentration in the bulk leads to non-monotonic behavior of the electrostatic potential profile. An increase in the co-solvent concentration in the bulk at its sufficiently small values makes the electrostatic potential profile longer-ranged. Oppositely, when the co-solvent concentration in the bulk exceeds some threshold value, its further increase leads to decrease in electrostatic potential at all distances from the electrode.
\end{abstract}

\maketitle
\section{Introduction}
The Poisson-Boltzmann (PB) equation is the simplest and very efficient tool for describing distribution of charged particles near the macroscopic charged objects in many areas, such as biophysics, electrochemistry, chemical engineering, etc \cite{Israelachvili}. As is well known, the PB equation is based on the mean-field theory that makes its application  to real systems quite problematic. Firstly, the mean-field theory itself does not allow us to take into account the effects of the ionic correlations that is crucial for medium and high concentrated electrolyte solutions. Secondly, considering the solvent as a continuous dielectric medium makes it impossible to study the effects of the solvent molecular structure. These two factors have motivated the researchers to improve the PB equation in the last two decades \cite{Podgornik_Review,Ben-Yaakov_Review}. At present, great efforts have been made to modify the PB equation with respect to ionic correlations \cite{Podgornik_1989,Netz,Moreira_Netz,Netz_Orland,Forsman,Bazant2011}, the dipole structure of the solvent \cite{Coalson_1996,Andelman_2007,Andelman_2012,Buyukdagli_2013,Buyukdagli_2014}, polarizability and permanent dipole of ions \cite{Frydel,Buyukdagli_2013} as well as their excluded volume \cite{Andelman_1997,Antypov_2005,Kornyshev,Maggs2015,Buyukdagli,Buyukdagli_2}, the dielectric decrements of ions \cite{Ben-Yaakov_2011,Andelman_2015,Hatlo}, and finally solvent quadrupolarizability \cite{Slavchov}.

Most of these researches are devoted to the influence of the different microscopic ionic parameters on the macroscopic quantities of the electric double layer, such as local concentration of ions on the electrode, disjoining pressure, and double layer differential capacitance. The latter is one of the most important quantities for the electrochemical applications. In a recent work \cite{Budkov} we showed in the framework of the field-theoretical approach that if an electrolyte solution is mixed with some strongly polarizable dielectric co-solvent, then the variation of the differential capacitance becomes the greater the stronger polarizability grows. We also demonstrated that in contrast to the co-solvent polarizability the permanent dipole of the co-solvent molecules only slightly affects the differential capacitance. Moreover, due to the fact that the above mentioned theory described the ions and co-solvent molecules as point particles, the effects of the excluded volume were fully ignored. However, as was clearly showed by Kornyshev \cite{Kornyshev} in the framework of lattice gas model, the excluded volume of ions must strongly affects the value of differential capacitance in the region of high surface potentials.

In this work we continue our theoretical research into the co-solvent polarizability influence on the double layer differential capacitance. We obtain the expression for the grand thermodynamic potential as a functional of electrostatic potential profile within the density functional approach on the level of local density approximation, taking into account the electrostatic interactions of ions and co-solvent molecules as well as their excluded volume. We derive the modified PB equation, considering the three-component symmetric lattice gas model as a reference system and minimizing the grand thermodynamic potential with respect to the electrostatic potential. We apply this equation to the theory of electric double layer, studying the behavior of differential capacitance and local co-solvent concentration on the electrode as the functions of surface potential as well as a behavior of electrostatic potential profile with varying the polarizability and concentration of co-solvent in the bulk solution.

\section{Theory}
\subsection{General formalism}
We consider an electrolyte solution containing $N_{+}$ ions carrying a charge $q_{+}>0$, $N_{-}$ ions carrying a charge $q_{-}<0$, and a solvent which we shall model as a continuous dielectric medium with dielectric permittivity $\varepsilon_{s}$. Moreover, we consider $N$ molecules of a co-solvent which have a polarizability $\alpha$. To describe the thermodynamic properties of such system, we shall use the variant of density functional theory at the level of local density approximation developed recently in the work \cite{Maggs2015}.

The grand thermodynamic potential of the electrolyte solution mixed with the polarizable co-solvent can be written as
\begin{equation}
\nonumber
\Omega=-\int \frac{\varepsilon(\bold{r})\left(\nabla{\psi}(\bold{r})\right)^2}{8\pi}  d\bold{r}+\int \rho_{c}(\bold{r})\psi(\bold{r})d\bold{r}
\end{equation}
\begin{equation}
\label{eq:Omega1}
+\int\left(f(c_{+}(\bold{r}),c_{-}(\bold{r}),n(\bold{r}))-\mu_{+}c_{+}(\bold{r})-\mu_{-}c_{-}(\bold{r})-\mu n(\bold{r})\right)d\bold{r} ,
\end{equation}
where $\varepsilon(\bold{r})=\varepsilon_{s}+4\pi \alpha n(\bold{r})$ is the local dielectric permittivity, $c_{\pm}(\bold{r})$ is the local concentrations of ions, $n(\bold{r})$ is the local concentration of co-solvent, $\rho_{c}(\bold{r})=q_{+}c_{+}(\bold{r})+q_{-}c_{-}(\bold{r})$ is the charge density, $f$ is the density of free energy of the reference system (see below).

Rewriting the grand thermodynamic potential (\ref{eq:Omega1}) as
\begin{equation}
\label{eq:Omega2}
\Omega=\int \left(-\frac{\varepsilon_{s}\left(\nabla{\psi}\right)^2}{8\pi}+f(c_{+},c_{-},n)-(\mu_{+}-q_{+}\psi)c_{+}-(\mu_{-}-q_{-}\psi)c_{-}-\left(\mu+\frac{\alpha}{2}\left(\nabla{\psi}\right)^2\right) n\right)d\bold{r},
\end{equation}
and using the thermodynamic relation for the pressure
\begin{equation}
P=c_{+}\mu_{+}+c_{-}\mu_{-}+n\mu- f,
\end{equation}
we eventually obtain
\begin{equation}
\label{eq:Omega3}
\Omega[\psi]=-\int \left(\frac{\varepsilon_{s}\left(\nabla{\psi}\right)^2}{8\pi}+ P(\mu_{+}-q_{+}\psi,\mu_{-}-q_{-}\psi,\mu+\frac{\alpha}{2}\left(\nabla{\psi}\right)^2)\right) d\bold{r}.
\end{equation}
Thus, if the explicit function $P=P(\mu_{+},\mu_{-},\mu)$ is known, one can obtain the explicit equation for the electrostatic potential $\psi(\bold{r})$ by minimizing the functional (\ref{eq:Omega3}).
To take into account the excluded volume of co-solvent and ions, we consider the lattice gas model (without the attractive Van-der-Waals interactions between the particles) as a reference system for which the explicit dependence $P=P(\mu_{+},\mu_{-},\mu)$ is well known:
\begin{equation}
\label{eq:P}
P=\frac{k_{B}T}{v}\ln\left(1+e^{\beta\mu_{+}}+e^{\beta\mu_{-}}+e^{\beta\mu}\right),
\end{equation}
where $v$ is the volume occupied by a particle of lattice gas, $T$ is the temperature, $k_{B}$ is the Boltzmann constant, $\beta=1/k_{B}T$.

Therefore, we obtain the following functional:
\begin{equation}
\label{eq:Omega4}
\Omega[\psi]=-\int \left(\frac{\varepsilon_{s}\left(\nabla{\psi}\right)^2}{8\pi}+\frac{k_{B}T}{v}\ln\left(1+e^{\beta\left(\mu_{+}-q_{+}\psi\right)}+e^{\beta\left(\mu_{-}-q_{-}\psi\right)}+e^{\beta\left(\mu+\frac{\alpha}{2}\left(\nabla{\psi}\right)^2\right)}\right)\right)d\bold{r}.
\end{equation}
Further, minimizing the functional (\ref{eq:Omega4}) and using the expressions for the chemical potentials of species
\begin{equation}
\label{eq:mu}
\mu_{\pm}=k_{B}T\ln{\frac{c_{\pm,b}v}{1-v(c_{+,b}+c_{-,b}+n_{b})}}, ~~\mu=k_{B}T\ln{\frac{n_{b}v}{1-v(c_{+,b}+c_{-,b}+n_{b})}},
\end{equation}
we arrive at the modified Poisson-Boltzmann equation with accounting for the polarizability of co-solvent molecules, their excluded volume, and the excluded volume of electrolyte ions
\begin{equation}
\label{eq:PBeq1}
\nabla(\varepsilon(\bold{r})\nabla{\psi}(\bold{r}))=-\frac{4\pi \left(q_{+}c_{+,b}e^{-\beta q_{+}\psi(\bold{r})}+q_{-}c_{-,b}e^{-\beta q_{-}\psi(\bold{r})}\right)}{1+v\left(c_{+,b}\left(e^{-\beta q_{+}\psi(\bold{r})}-1\right)+c_{-,b}\left(e^{-\beta q_{-}\psi(\bold{r})}-1\right)+n_{b}\left(e^{\frac{\beta\alpha}{2}\left(\nabla{\psi}(\bold{r})\right)^2}-1\right)\right)},
\end{equation}
where $c_{\pm,b}$ is the bulk concentrations of ions, $n_{b}$ is the bulk co-solvent concentration;
\begin{equation}
\label{eq:perm1}
\varepsilon(\bold{r})=\varepsilon_{s}+\frac{4\pi\alpha n_{b}  e^{\frac{\beta\alpha}{2}\left(\nabla{\psi}(\bold{r})\right)^2}}{1+v\left(c_{+,b}\left(e^{-\beta q_{+}\psi(\bold{r})}-1\right)+c_{-,b}\left(e^{-\beta q_{-}\psi(\bold{r})}-1\right)+n_{b}\left(e^{\frac{\beta\alpha}{2}\left(\nabla{\psi}(\bold{r})\right)^2}-1\right)\right)}
\end{equation}
is the local dielectric permittivity of the electrolyte solution. When there are no co-solvent molecules in the electrolyte solution ($n_{b}=0$), we arrive at the equation obtained firstly by Borukhov et al \cite{Andelman_1997} and Kornyshev \cite{Kornyshev}
\begin{equation}
\varepsilon_{s}\nabla^2\psi(\bold{r})=-\frac{4\pi\left(q_{+}c_{+,b}e^{-\beta q_{+}\psi(\bold{r})}+q_{-}c_{-,b}e^{-\beta q_{-}\psi(\bold{r})}\right)}{1+v\left(c_{+,b}\left(e^{-\beta q_{+}\psi(\bold{r})}-1\right)+c_{-,b}\left(e^{-\beta q_{-}\psi(\bold{r})}-1\right)\right)}.
\end{equation}

In limit of the point particles (when $v\rightarrow 0$) equation (\ref{eq:PBeq1}) looks as follows
\begin{equation}
\label{eq:PBeq2}
\nabla(\varepsilon(\bold{r})\nabla{\psi}(\bold{r}))=-4\pi \left(q_{+}c_{+,b}e^{-\beta q_{+}\psi(\bold{r})}+q_{-}c_{-,b}e^{-\beta q_{-}\psi(\bold{r})}\right),
\end{equation}
where $\varepsilon(\bold{r})=\varepsilon_{s}+4\pi\alpha n_{b}e^{\frac{\beta\alpha}{2}\left(\nabla{\psi}(\bold{r})\right)^2}$ is the local dielectric permittivity in the approximation of point particles.  It should be noted that equation (\ref{eq:PBeq2}) was obtained in the recent work \cite{Budkov} within the field-theoretical approach.

\subsection{Theory of electric double layer}

As an application of the modified PB equation (\ref{eq:PBeq1}-\ref{eq:perm1}), we formulate the generalized Kornyshev's theory \cite{Kornyshev,Barrat_Hansen}. We consider a system containing a charged electrode, which we shall model as a charged flat surface with a  surface charge density $\sigma$, the ions of 1:1 electrolyte (i.e. when $q_{+}=-q_{-}=e$; $e$ is the elementary charge), and the molecules of the polarizable co-solvent with a polarizability $\alpha$. In this case the average concentrations of ions in the bulk are equal, i.e., $c_{+,b}=c_{-,b}=c$.
Choosing $z$ axis perpendicular to the electrode and placing the origin on it, one can write the grand thermodynamic  potential per unit area of the electrode as follows:
\begin{equation}
\label{eq:Omega5}
\Omega[\psi]=-\int\limits_{0}^{\infty}dz \left(\frac{\varepsilon_{s}\left(\psi^{\prime}(z)\right)^2}{8\pi}+\frac{k_{B}T}{v}\ln\left(1+e^{\beta\left(\mu_{+}-e\psi(z)\right)}+e^{\beta\left(\mu_{-}+e\psi(z)\right)}+e^{\beta\left(\mu+\frac{\alpha}{2}\left(\psi^{\prime}(z)\right)^2\right)}\right)\right),
\end{equation}

Since the integrand in (\ref{eq:Omega5}) does not depend on coordinate $z$ explicitly, the Euler-Lagrange equation has a first integral which determines the condition of the solution mechanical equilibrium
\begin{equation}
\label{eq:press}
P\left(\mu_{+}-e\psi,\mu_{-}+e\psi,\mu+\frac{\alpha\mathcal{E}^2}{2}\right)-\frac{\varepsilon_{s}\mathcal{E}^2}{8\pi}-\alpha\mathcal{E}^2 n\left(\mu_{+}-e\psi,\mu_{-}+e\psi,\mu+\frac{\alpha\mathcal{E}^2}{2}\right)=P\left(\mu_{+},\mu_{-},\mu\right),
\end{equation}
where the local electric field $\mathcal{E}(z)=-\psi^{\prime}(z)$ and the local co-solvent concentration $n=\partial{P}/\partial{\mu}$ are introduced. The first term in the left-hand side of eq. (\ref{eq:press}) determines the pressure which is related to the excluded volume of particles, whereas the second and third terms determine the so-called disjoining pressure contribution which is due to the electrostatic interactions \cite{Barrat_Hansen}.

Further, substituting the expressions for the bulk chemical potentials of species (\ref{eq:mu}) and for the pressure (\ref{eq:P}) into the equation (\ref{eq:press}), we eventually obtain
\begin{equation}
\label{eq:E1}
1+\left(2c\left(\cosh(\beta e\psi(z))-1\right)+n_{b}\left(e^{\frac{\beta\alpha \mathcal{E}^2(z)}{2}}-1\right)\right)v=e^{\frac{v}{k_{B}T}\left(\frac{\varepsilon_{s}\mathcal{E}^2(z)}{8\pi}+\frac{\alpha n_{b}\mathcal{E}^2(z)e^{\frac{\beta\alpha\mathcal{E}^2(z)}{2}}}{1+\left(2c\left(\cosh{\beta e\psi(z)}-1\right)+n_{b}\left(e^{\frac{\beta\alpha\mathcal{E}^2(z)}{2}}-1\right)\right)v}\right)}.
\end{equation}

In the limit $v\rightarrow 0$ we obtain the following equation
\begin{equation}
\label{eq:E2}
\frac{\varepsilon_{s}\mathcal{E}^2(z)}{8\pi}+n_{b} k_{B}T\left(1-e^{\frac{\beta\alpha \mathcal{E}^2(z)}{2}}\right)+n_{b}\alpha \mathcal{E}^2(z)e^{\frac{\beta\alpha \mathcal{E}^2(z)}{2}}=2c k_{B}T\left(\cosh{\beta e\psi(z)}-1\right)
\end{equation}
which was first obtained in the work \cite{Budkov}.

To obtain the potential profile $\psi(z)$, we should first solve the eq. (\ref{eq:E1}) as a transcendental equation numerically (for instance, by Newton's method) with respect to $\mathcal{E}=-\psi^{\prime}(z)$ at different values of $\psi$.  Thus, we obtain the function $\mathcal{E}=\mathcal{E}(\psi)$. In order to obtain the potential profile $\psi(z)$, we solve numerically the equation $\psi^{\prime}=-\mathcal{E}(\psi)$ with use of the standard boundary condition
\begin{eqnarray}
\label{eq:boundary_cond1}
-\varepsilon(0)\psi^{\prime}(0)=4\pi\sigma,
\end{eqnarray}
where the local dielectric permittivity of the electrolyte solution
\begin{equation}
\varepsilon(z)=\varepsilon_{s}+4\pi\alpha n(z)
\end{equation}
is introduced. The local co-solvent concentration can be expressed as follows
\begin{equation}
n(z)=\frac{n_{b}e^{\frac{\beta\alpha \mathcal{E}^2(z)}{2}}}{1+v\left(2c\left(\cosh{\beta e\psi(z)}-1\right)+n_{b}\left(e^{\frac{\beta\alpha \mathcal{E}^2(z)}{2}}-1\right)\right)}.
\end{equation}

To calculate the differential capacitance $C=\partial{\sigma}/\partial{\psi_{0}}$ as a function of the surface electrostatic potential $\psi_{0}=\psi(0)$ which is usually an experimentally controllable parameter, we should calculate the surface charge density $\sigma$. For this purpose we use the first integral (\ref{eq:E1}) written for $z=0$
\begin{equation}
\label{eq:E3}
1+\left(2c\left(\cosh(\beta e\psi_{0})-1\right)+n_{b}\left(e^{\frac{\beta\alpha \mathcal{E}_{0}^2}{2}}-1\right)\right)v=e^{\frac{v}{k_{B}T}\left(\frac{\varepsilon_{s}\mathcal{E}_{0}^2}{8\pi}+\frac{\alpha n_{b}\mathcal{E}_{0}^2e^{\frac{\beta\alpha\mathcal{E}_{0}^2}{2}}}{1+\left(2c\left(\cosh{\beta e\psi_{0}}-1\right)+n_{b}\left(e^{\frac{\beta\alpha\mathcal{E}_{0}^2}{2}}-1\right)\right)v}\right)},
\end{equation}
where $\mathcal{E}_{0}=\mathcal{E}(0)$,
and the boundary condition  (\ref{eq:boundary_cond1}) may be rewritten in the form
\begin{equation}
\label{eq:boundary_cond2}
\sigma=\frac{1}{4\pi}\left(\varepsilon_{s}+\frac{4\pi\alpha n_{b}  e^{\frac{\beta\alpha\mathcal{E}_{0}^2}{2}}}{1+v\left(2c\left(\cosh{\beta e\psi_{0}}-1\right)+n_{b}\left(e^{\frac{\beta\alpha \mathcal{E}_{0}^2}{2}}-1\right)\right)}\right)\mathcal{E}_{0}.
\end{equation}

Solving the system of coupled nonlinear equations (\ref{eq:E3}-\ref{eq:boundary_cond2}) numerically with respect to $\mathcal{E}_{0}$ and $\sigma$ at different values of $\psi_{0}$, we obtain the dependence $\sigma=\sigma(\psi_{0})$ that allows us to obtain the differential capacitance profile $C=C(\psi_{0})$ (see the next section).

\section{Numerical results and discussion}
Turning to the numerical calculations, we determine the following reduced parameters: $\tilde{n}_{b}=n_{b}v$, $\tilde{\alpha}=\alpha/v\varepsilon_{s}$, $\tilde{\mathcal{E}}=\beta ev^{1/3}\mathcal{E}$, $u=\beta e\psi$, $\tilde{z}=z/v^{1/3}$, and $\tilde{\sigma}=\sigma\beta ev^{1/3}/\varepsilon_{s}$. We first discuss the behavior of the differential capacitance as the function of surface potential. The reduced differential capacitance $\tilde{C}=C/v^{1/3}\varepsilon_{s}$ can be calculated as
\begin{equation}
\tilde{C}=\frac{\partial{\tilde{\sigma}}}{\partial{u}_{0}},
\end{equation}
where $u_{0}=u(0)$. The system of coupled equations (\ref{eq:E3}-\ref{eq:boundary_cond2}) can be rewritten in the dimensionless form as
\begin{equation}
\label{eq:E4}
1+2\tilde{c}\left(\cosh{u_{0}}-1\right)+\tilde{n}_{b}\left(e^{\frac{\tilde{\alpha}\tilde{\mathcal{E}}_{0}^2}{2\xi}}-1\right)=e^{\frac{1}{\xi}\left(\frac{\tilde{\mathcal{E}}_{0}^2}{8\pi}+\frac{\tilde{\alpha} \tilde{n}_{b}\tilde{\mathcal{E}}_{0}^2e^{\frac{\tilde{\alpha}\tilde{\mathcal{E}}_{0}^2}{2\xi}}}{1+2\tilde{c}\left(\cosh{u_{0}}-1\right)+\tilde{n}_{b}\left(e^{\frac{\tilde{\alpha}\tilde{\mathcal{E}}_{0}^2}{2\xi}}-1\right)}\right)},
\end{equation}
and
\begin{equation}
\label{eq:boundary_cond3}
\tilde{\sigma}=\frac{1}{4\pi}\left(1+\frac{4\pi\tilde{\alpha} \tilde{n}_{b}  e^{\frac{\tilde{\alpha}\tilde{\mathcal{E}}_{0}^2}{2\xi}}}{1+2\tilde{c}\left(\cosh{u_{0}}-1\right)+\tilde{n}_{b}\left(e^{\frac{\tilde{\alpha} \tilde{\mathcal{E}}_{0}^2}{2\xi}}-1\right)}\right)\tilde{\mathcal{E}}_{0},
\end{equation}
where $\xi=l_{B}/v^{1/3}$, $l_B=e^2/{\varepsilon_{s}k_{B}T}$ is the Bjerrum length.

The first integral (\ref{eq:E1}) of the modified Poisson-Boltzmann equation can be also rewritten in the dimensionless form as follows
\begin{equation}
\label{eq:E5}
1+2\tilde{c}\left(\cosh{u(\tilde{z})}-1\right)+\tilde{n}_{b}\left(e^{\frac{\tilde{\alpha}\tilde{\mathcal{E}}^2(\tilde{z})}{2\xi}}-1\right)=e^{\frac{1}{\xi}\left(\frac{\tilde{\mathcal{E}}^2(\tilde{z})}{8\pi}+\frac{\tilde{\alpha} \tilde{n}_{b}\tilde{\mathcal{E}}^2(\tilde{z})e^{\frac{\tilde{\alpha}\tilde{\mathcal{E}}^2(\tilde{z})}{2\xi}}}{1+2\tilde{c}\left(\cosh{u(\tilde{z})}-1\right)+\tilde{n}_{b}\left(e^{\frac{\tilde{\alpha}\tilde{\mathcal{E}}^2(\tilde{z})}{2\xi}}-1\right)}\right)}.
\end{equation}

We use the following values of the physical parameters $\varepsilon_{s}=80$, $T=300~K$, $c=0.1~mol/L$ $v^{1/3}=0.3~nm$ which yield a set of the reduced parameters: $\tilde{c}=1.63\times 10^{-3}$, $\xi=2.32$. Fig. 1a demonstrates the differential capacitance profiles $\tilde{C}=\tilde{C}(u_{0})$ for the small co-solvent polarizabilities and the fixed bulk co-solvent concentration $\tilde{n}_{b}=0.5$. As is seen, increasing the co-solvent polarizability may generate a differential capacitance enhancement in the region of surface potentials less than the 'saturation' potential $u_{sat}$ (a surface potential at which the maximum of the differential capacitance is achieved). However, if the surface potential is in the region of electric double layer saturation, increase in the co-solvent polarizability provokes a decrease in the differential capacitance (see Fig. 1a). Oppositely, when the co-solvent polarizability exceeds some critical value, its increase leads to a different behavior of the differential capacitance. Namely, increasing the co-solvent polarizability in this case generates a significant growth of the differential capacitance in the wide range of surface potentials (see Fig. 1b). In order to understand these two qualitatively different regimes, let us consider the behavior of co-solvent concentration on the electrode $\tilde{n}_{s}=\tilde{n}(0)$ as the function of surface potential $u_{0}$ at different co-solvent polarizabilities $\tilde{\alpha}$. Fig. 2 demonstrates the values of $\tilde{n}_{s}$ as the functions of surface potential at different co-solvent polarizabilities. As one can see, at sufficiently small co-solvent polarizability the cosolvent molecules are depleted at the electrode. On the contrary, when the co-solvent polarizability exceeds some threshold value, the co-solvent molecules create an adsorption layer on the charged electrode. These two regimes are clearly demonstrated by fig. 3, where the co-solvent concentration profiles $\tilde{n}(\tilde{z})$ are depicted. Thus, two different regimes of the differential capacitance behavior are related to the depletion and adsorption of co-solvent molecules at the charged electrode.

Figures 4a,b show the differential capacitance profiles at different values of the dimensionless co-solvent concentration $\tilde{n}_{b}$ at the fixed co-solvent polarizability $\tilde{\alpha}=0.3$. As one can see, an increase in the co-solvent concentration in the bulk solution can shift significantly the maximum of differential capacitance by two qualitatively different manners. Namely, at the sufficiently small co-solvent concentration its increase leads to a shift of the differential capacitance maximum to the region of higher surface potentials (see fig. 4a). It means that an additive of the small quantity of the polarizable co-solvent to the electrolyte solution prevents the saturation of the electric double layer. In the case, when the co-solvent concentration exceeds the threshold value, the maximum of differential capacitance shifts to the region of lower surface potentials (see fig.4b). Figure 5 shows the dependencies of the saturation potential $u_{sat}$ on the co-solvent concentration $\tilde{n}_{b}$ at different values of co-solvent polarizability $\tilde{\alpha}$. As it is shown, the non-monotonic behavior of the saturation potential with varying co-solvent concentration occurs at sufficiently large co-solvent polarizability only. However, an additive of the co-solvent with sufficiently small polarizability leads to the shift of the saturation potential to lower potentials for all the considered co-solvent concentrations. It should be noted that non-monotonic behavior of the saturation potential with increasing co-solvent concentration can be of interest to electrochemical applications, where it is necessary to control the differential capacitance.

Fig. 6 demonstrates the comparison between the differential capacitance profiles obtained by the present theory and our previous theory. As one can see, previous theory is valid at small surface potentials only. Indeed, accounting for the excluded volume of both ions and molecules of the co-solvent results in a decrease in the differential capacitance in the region of high surface potential compared to the simplified theory of point particles. The latter means that the dramatic increase in the differential capacitance at high surface potentials predicted in work \cite{Budkov} is unphysical.

Finally, we discuss the influence of the co-solvent concentration and co-solvent polarizability on the electrostatic potential profile $u(\tilde{z})$. As well as in our previous theory, an increase in the co-solvent polarizability leads to longer-ranged electrostatic potential profiles (Fig. 7). The latter is due to the fact that an increase in this variable results in higher local dielectric permittivity that, in turn, leads to a decrease in the electrode charge screening. However, an increase in the bulk co-solvent concentration leads to more complex behavior of the electrostatic potential profile. Namely, increasing the bulk co-solvent concentration at its sufficiently small values makes the electrostatic potential profile longer-ranged. Nevertheless, when the co-solvent concentration in the bulk exceeds some threshold value, its further increase leads to a decrease in the electrostatic potential at all distances from the electrode (Fig. 8).  Such behavior of the electrostatic potential depending on the bulk co-solvent concentration is different on that predicted by our previous simplified theory of point particles, where the potential profile becomes longer-ranged at all co-solvent concentartions.

\section{Conclusion}
In this work based on the density functional formalism on the level of local density approximation, we have developed a modified Poisson-Boltzmann equation with an explicit account of the polarizable co-solvent in combination with the excluded volume of ions and co-solvent molecules. We have applied the modified Poisson-Boltzmann equation to electric double layer theory and shown that like in our previous simplified theory \cite{Budkov} (where all particles of the electrolyte solution were considered as point ones), the present theory predicts the influence of the co-solvent polarizability on the differential capacitance. Namely, in the case of small co-solvent polarizabilities under sufficiently small surface potentials of electrode the differential capacitance grows significantly with increasing of the co-solvent polarizability as well as bulk co-solvent concentration. Oppositely, when the surface potential exceeds some threshold value (which is close to the saturation potential), the growth of the co-solvent polarizability and bulk co-solvent concentration results in decrease in the differential capacitance. However, when the co-solvent polarizability exceeds some threshold value, its increase generates a considerable growth of the differential capacitance in the region of the double layer saturation. We have established that two qualitatively different regimes of the differential capacitance behavior are caused by the depletion and adsorption of co-solvent molecules at the charged electrode. We have also shown that an additive of the sufficiently strong polarizable co-solvent to an electrolyte solution can significantly shift the maximum of differential capacitance by two qualitatively different ways. Namely, a small additive of co-solvent results in the shift of differential capacitance maximum to the higher surface potentials. However, when the bulk co-solvent concentration exceeds the threshold value, the maximum of differential capacitance shifts to the lower surface potentials. We have shown that increase in the co-solvent polarizability results in longer-ranged electrostatic potential profile. Finally, we have obtained that at sufficiently small co-solvent concentration in the bulk its increase makes the electrostatic potential profile longer-ranged. Nevertheless, when the co-solvent concentration in the bulk exceeds some threshold value, its further increase leads to a decrease in electrostatic potential.

Now we would like to discuss the limitations of the present theory. It is well known, the lattice gas model highly underestimates the pressure in the bulk at high number densities of particles for the off-lattice hard spheres system \cite{Sanchez1976}. Moreover, the lattice gas model highly overestimates the differential capacitance obtained by MD computer simulations in the wide range of surface potential \cite{Fedorov2008}. That is why the lattice gas model cannot be used for quantitative predictions of both thermodynamic and electrochemical variables, but only for their qualitative evaluations. To get more reliable quantitative results, one can use more precise Percus-Yevick or Carnahan-Starling equations of state. However, the application of these equations of state will involve more difficult numerical calculations \cite{Maggs2015}. The next limitation is related to the fact that the present theory is based on the local density approximation and fully ignores the nonlocal packing effects which have a short-range nature and must be important for the ions and co-solvent molecules near the electrode \cite{Howard2010,Bazant2011}. However, we believe that such short-ranged effects could not drastically affect the double layer differential capacitance which should be determined mostly by the long-range correlations of particles. On the other hand, the effects of co-solvent polarizability related to the long-range correlations of particles \cite{Schroder_2015,Cavalcante_2014} should be qualitatively described on the level of mean-field approximation. Unfortunately, we cannot give an {\sl a priori} estimate within this formalism of the results obtained. The latter requires calculations based on the nonlocal density functional theory or computer simulations. In the present theory, we have considered the solvent as continuous dielectric medium with fixed dielectric permittivity. In other words, we have assumed that the solvent dielectric permittivity near the charged electrode is the same as that in the bulk solution. However, as is well known, such assumption cannot be correct for the sufficiently large surface charge density of the electrode. Indeed, the application of sufficiently large electric field can lead to significant decrease of the water dielectric permittivity \cite{Booth1951,Sutman1998,Yeh1999,Gongadze2015}.  That is why our theory gives highly overestimated polarizabilities of the co-solvent molecules $\alpha\simeq 200~\textup{\AA}^3$ ($\tilde{\alpha}\simeq 0.1$) for which the discussed phenomena might be realized. We believe that accounting for the effect of dielectric permittivity renormalization near the charged electrode might reduce the polarizability to the physically reasonable values ($\alpha\simeq 10~\textup{\AA}^3$). Nevertheless, we hope that our self-consistent field theory may be of use for qualitative evaluations in various electrochemical applications. Finally, it is worth noting that the present theory makes sense only in the case when the co-solvent polarizability significantly greater than the polarizability of solvent. Indeed, only in such case the consideration of the solvent as a continuous dielectric medium at a sufficient distance from the electrode may be justified. Evidently, this condition can be satisfied for the aromatic compounds dissolved in some aqueous electrolyte solution.

In conclusion, we would like to speculate on the possible application of our theory to the experimental systems. In our opinion, it can be applied to the theoretical description of the aromatic compounds solubilization in aqueous micellar solutions of amphiphilic imidazolium ionic liquids \cite{Luczak2013}.

\begin{acknowledgments}
We thank N. Georgi, A.I. Victorov and E.A. Safonova for fruitful discussions. We thank Reviewers for valuable comments that helped us to improve this  work. This research was supported by grant from the President of the Russian Federation (No MK-2823.2015.3).
\end{acknowledgments}

\newpage

\begin{figure}
\centerline{\includegraphics[scale = 1.2]{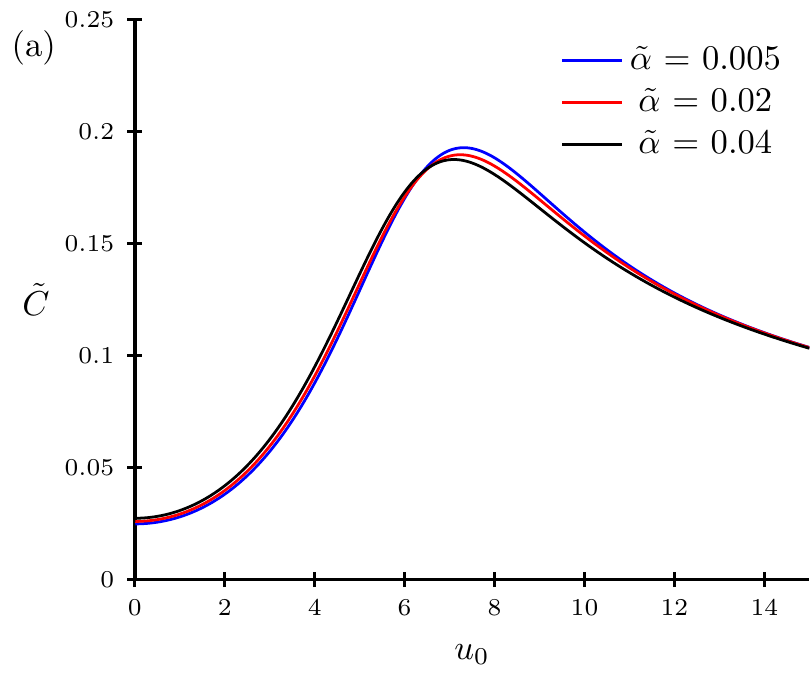}}
\centerline{\includegraphics[scale = 1.2]{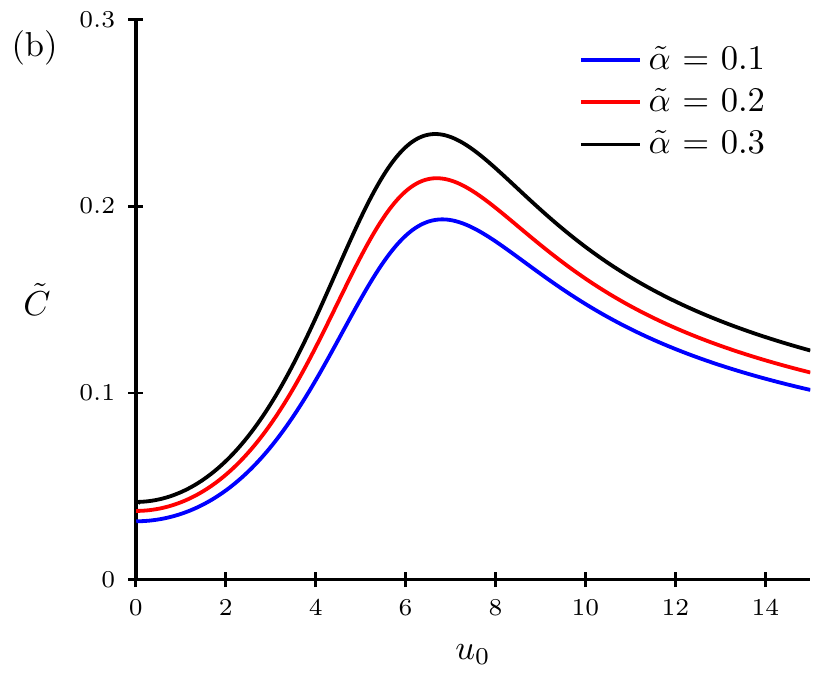}}
\caption{The differential capacitance profiles $\tilde{C}=\tilde{C}(u_{0})$ at different co-solvent polarizabilities: (a) $\tilde{\alpha}=0.005,~0.02,~0.04$ and (b) $\tilde{\alpha}=0.1,~0.2,~0.3$. The data are shown for $\tilde{c}=1.626\times 10^{-3}$, $\tilde{n}_{b}=0.5$, $\xi=2.32$.}
\label{fig1}
\end{figure}

\begin{figure}
\centerline{\includegraphics[scale = 1.2]{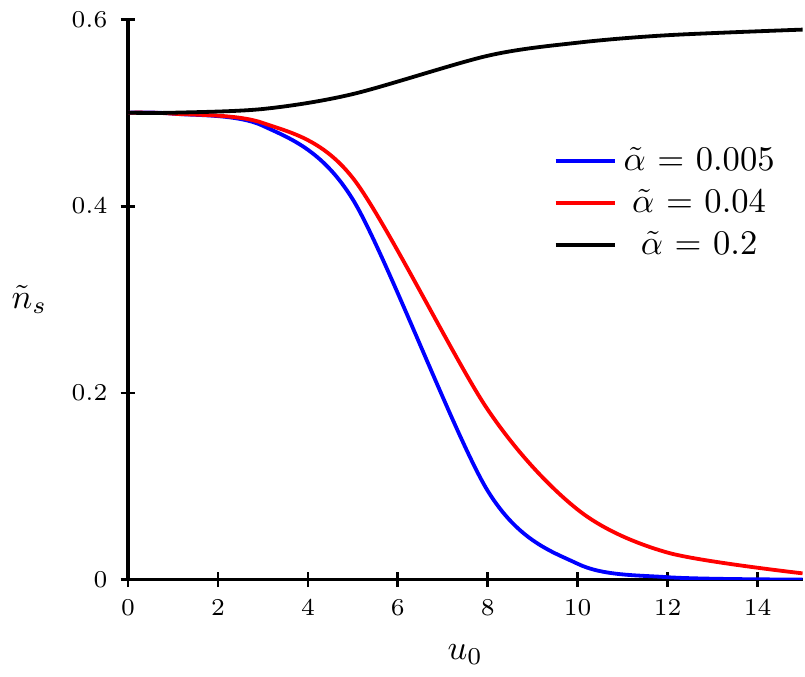}}
\caption{The local co-solvent concentration $\tilde{n}_{s}=\tilde{n}(0)$ on the electrode as a function of the surface potential $u_{0}$ at different co-solvent polarizabilities $\tilde{\alpha}=0.005,~0.04,~0.2$ and fixed bulk cosolvent concentration $\tilde{n}_{b}=0.5$. At sufficiently small co-solvent polarizability the co-solvent molecules are depleted near the strongly charged electrode. On the contrary, when the co-solvent polarizability exceeds some threshold value, the adsorption of the co-solvent molecules on the charged electrode takes place.    The data are shown for $\tilde{c}=1.626\times 10^{-3}$, $\xi=2.32$.}
\label{fig2}
\end{figure}

\begin{figure}
\centerline{\includegraphics[scale = 1.2]{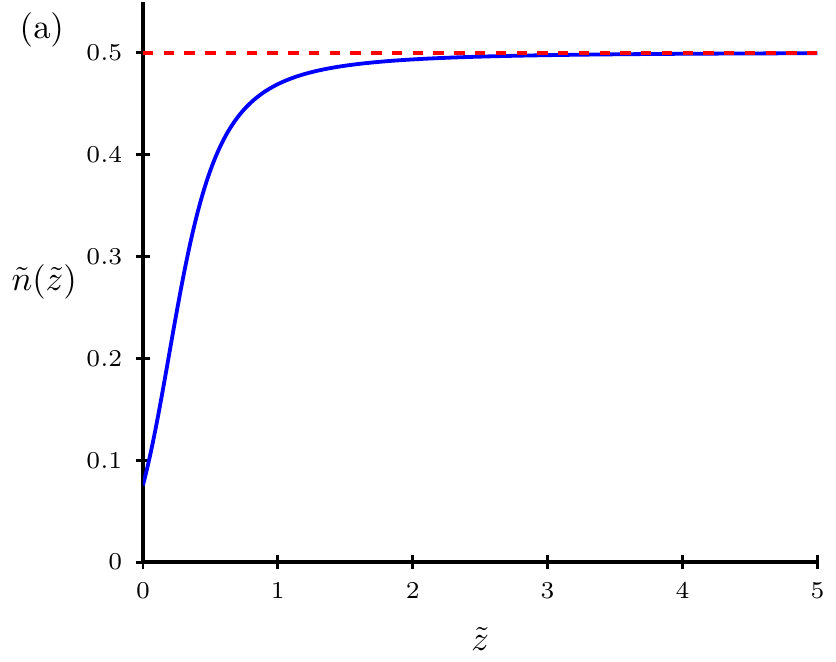}}
\centerline{\includegraphics[scale = 1.2]{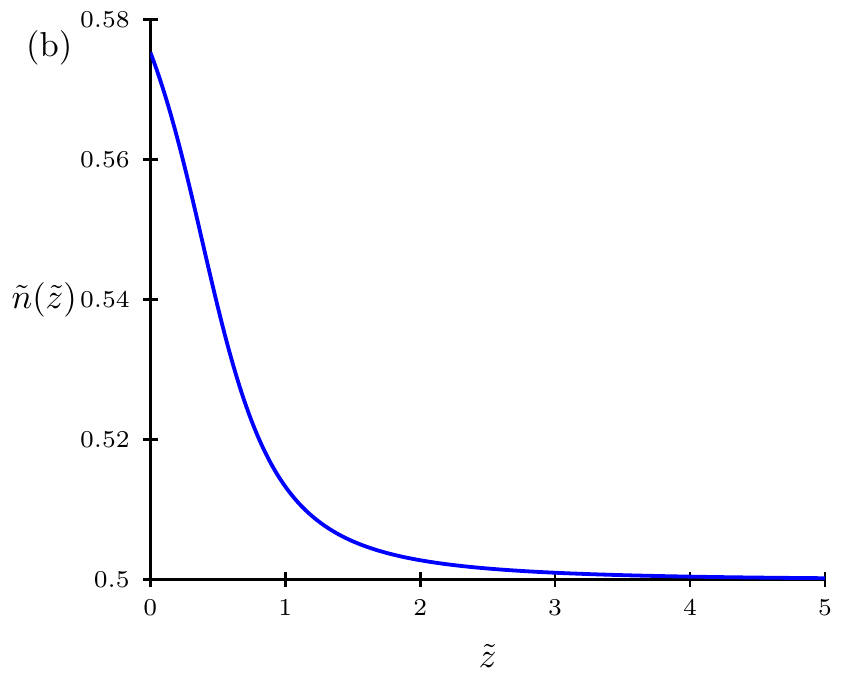}}
\caption{The co-solvent concentration profiles $\tilde{n}=\tilde{n}(\tilde{z})$ for (a) depletion and (b) adsorption regimes. The data are shown for (a) $\tilde{c}=1.626\times 10^{-3}$, $\tilde{\alpha}=0.04$, $\tilde{n}_{b}=0.5$, $\xi=2.32$ and (b) $\tilde{c}=1.626\times 10^{-3}$, $\tilde{\alpha}=0.2$, $\tilde{n}_{b}=0.5$, $\xi=2.32$.}
\label{fig3}
\end{figure}

\begin{figure}
\centerline{\includegraphics[scale = 1.2]{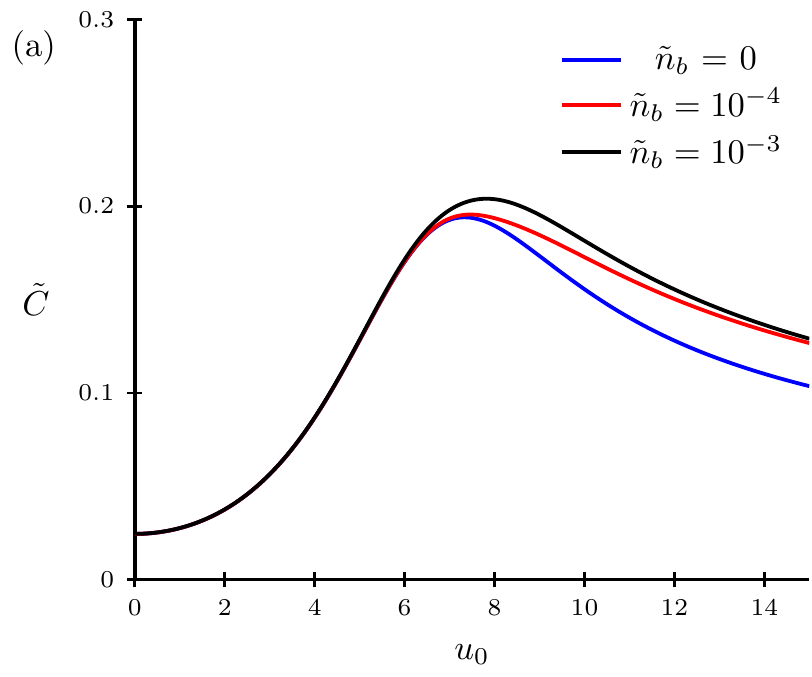}}
\centerline{\includegraphics[scale = 1.2]{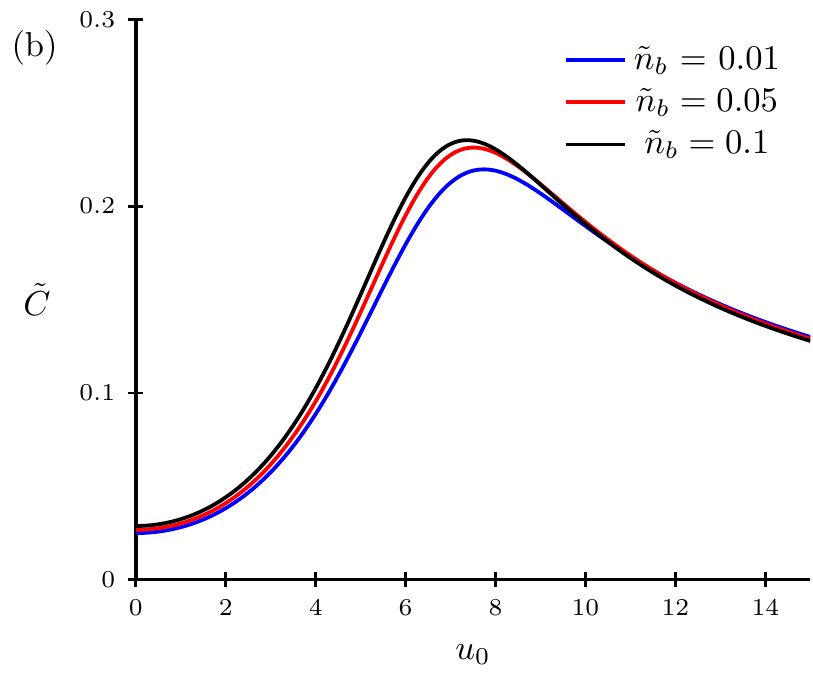}}
\caption{The differential capacitance profiles $\tilde{C}=\tilde{C}(u_{0})$ at different bulk co-solvent concentrations: (a) $\tilde{n}_{b}=0,10^{-4},10^{-3}$ and (b) $\tilde{n}_{b}=0.01,0.05,0.1$. At sufficiently small co-solvent concentartion its increase results in the shift of saturation potential to the region of higher potentials. When the co-solvent concentration exceeds some threshold value, the maximum of differential capacitance shifts to the region of lower potentials. The data are shown for $\tilde{c}=1.626\times 10^{-3}$, $\xi=2.32$, $\tilde{\alpha}=0.3$.}
\label{fig4}
\end{figure}

\begin{figure}
\centerline{\includegraphics[scale = 1.2]{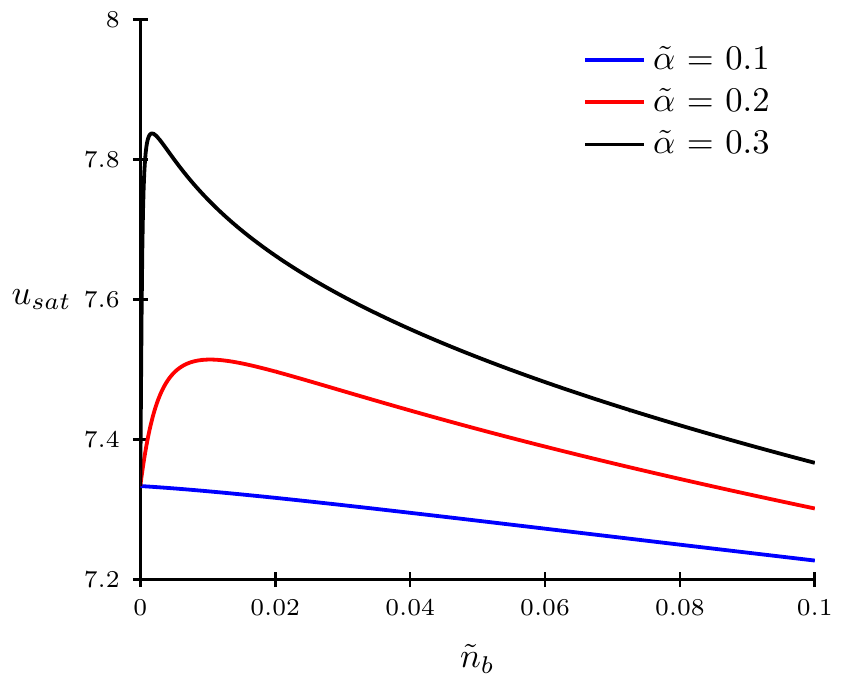}}
\caption{The dependencies of saturation potential $u_{sat}$ on the co-solvent concentration $\tilde{n}_{b}$ at the different co-solvent polarizabilities $\tilde{\alpha}=0.1,~0.2,~0.3$. The data are shown for $\tilde{c}=1.626\times 10^{-3}$, $\tilde{n}_{b}=0.5$, $\xi=2.32$.}
\label{fig5}
\end{figure}

\begin{figure}
\centerline{\includegraphics[scale = 1.2]{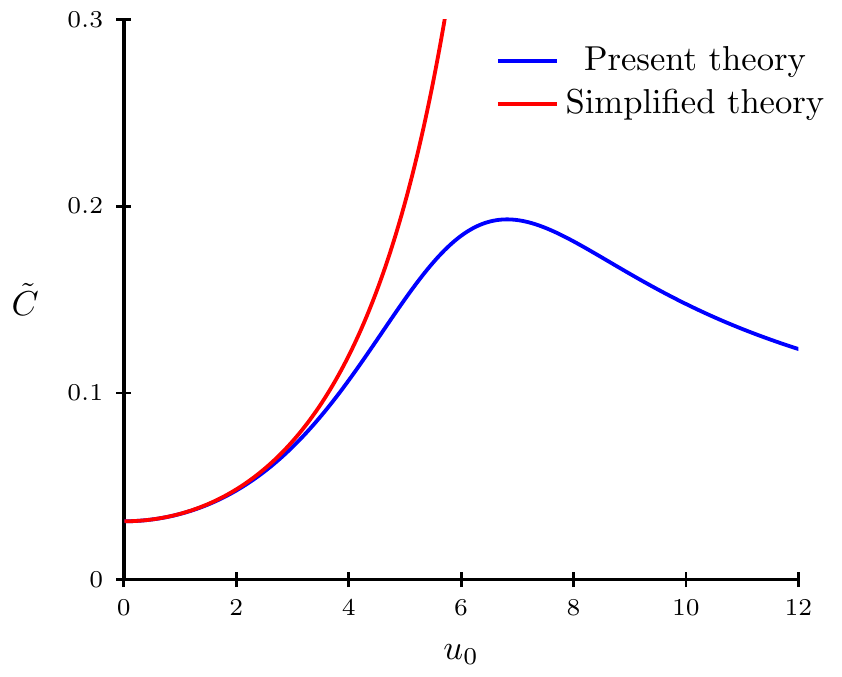}}
\caption{Comparison between differential capacitance profiles obtained by simplified theory and present theory. The simplified theory gives unphysical dramatic growth of the differential capacitance at the large surface potentials. The data are shown for $\tilde{c}=1.626\times 10^{-3}$, $\tilde{n}_{b}=0.5$, $\xi=2.32$, and $\tilde{\alpha}=0.1$.}
\label{fig6}
\end{figure}

\begin{figure}
\centerline{\includegraphics[scale = 1.2]{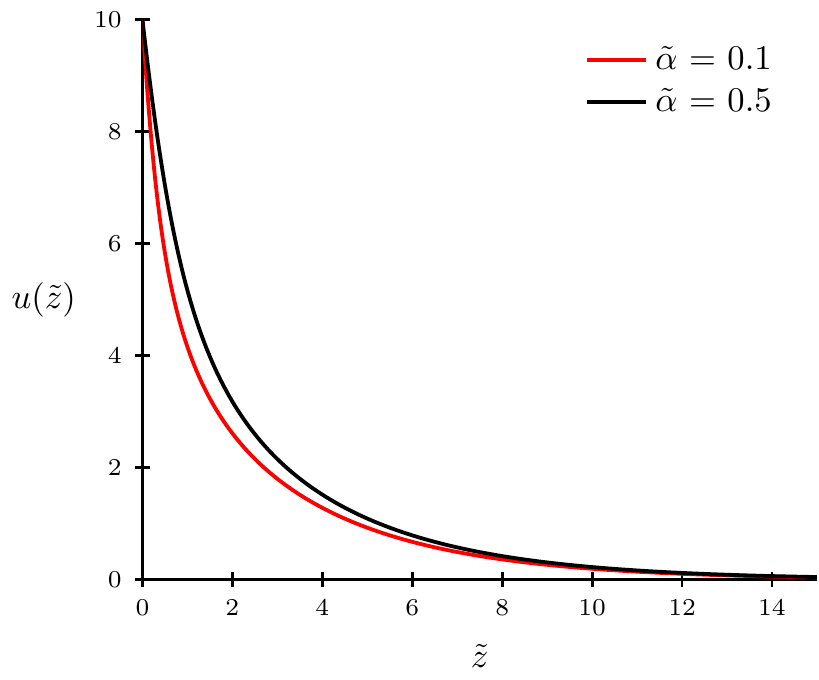}}
\caption{The electrostatic potential profiles $u=u(\tilde{z})$ at the different co-solvent polarizabilities $\tilde{\alpha}=0.1,~0.5$ and fixed co-solvent bulk concentration $\tilde{n}_{b}=0.5$. The potential profiles become more long-ranged at increase of co-solvent polarizability. The data are shown for $\tilde{c}=1.626\times 10^{-3}$, $\xi=2.32$, and $u_{0}=10$.}
\label{fig7}
\end{figure}

\begin{figure}
\centerline{\includegraphics[scale = 1.2]{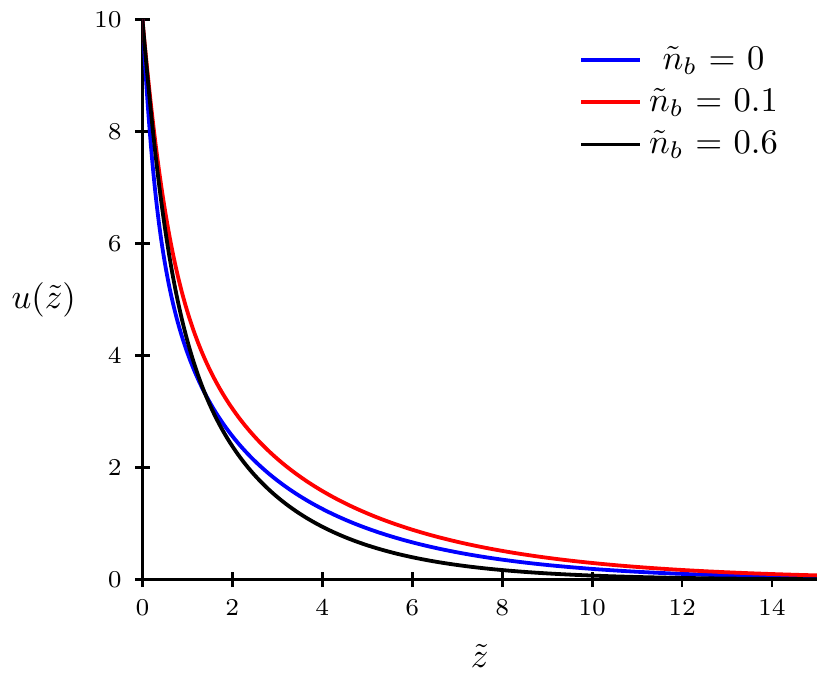}}
\caption{The electrostatic potential profiles $u=u(\tilde{z})$ at the different bulk co-solvent concentrations  $\tilde{n}_{b}=0,~0.1,~0.6$ and fixed co-solvent polarizability $\tilde{\alpha}=0.3$. The data are shown for $\tilde{c}=1.626\times 10^{-3}$, $\xi=2.32$, and $u_{0}=10$.}
\label{fig7}
\end{figure}


\begin{thebibliography}{99}
\bibitem{Israelachvili}
{Jacob N. Israelachvili} {\sl Intermolecular and surface forces} (Academic Press, 2011).

\bibitem{Podgornik_Review}
{Naji A., Kanduc M., Forsman J., Podgornik R.} J. Chem. Phys. $\bold{139}$, 150901 (2013).

\bibitem{Ben-Yaakov_Review}
{Dan Ben-Yaakov, David Andelman, Daniel Harries and Rudi Podgornik} J.Phys.: Condens. Matter. $\bold{21}$, 424106 (2009).

\bibitem{Netz}
{Netz R.R.} Eur. Phys. J. E $\bold{5}$, 557 (2001).

\bibitem{Moreira_Netz}
{A. G. Moreira and R. R. Netz} EPL $\bold{52}$, 705 (2000).

\bibitem{Podgornik_1989}
{Rudi Podgornik} J. Chem. Phys. $\bold{91}$, 5840 (1989).

\bibitem{Netz_Orland}
{R.R. Netz, H. Orland} The European Physical Journal E $\bold{2-3}$, 203 (2000).

\bibitem{Forsman}
{Forsman J.} J. Phys. Chem. B $\bold{108}$, 9236 (2004).

\bibitem{Bazant2011}
{Martin Z. Bazant, Brian D. Storey, and Alexei A. Kornyshev} PRL $\bold{106}$, 046102 (2011).

\bibitem{Coalson_1996}
{Rob D. Coalson, A. Duncan and N. B. Tal} J. Phys. Chem. B $\bold{100}$, 2612 (1996).

\bibitem{Andelman_2007}
{Abrashkin A., Andelman D., Orland H.} PRL $\bold{99}$, 077801 (2007).

\bibitem{Andelman_2012}
{Levy A., Andelman D., Orland H.} PRL $\bold{108}$, 227801 (2012).

\bibitem{Buyukdagli_2014}
{Sahin Buyukdagli and Ralf Blossey} J. Chem. Phys. $\bold{140}$, 234903 (2014).

\bibitem{Frydel}
{Frydel D.} J. Chem. Phys. $\bold{134}$, 234704 (2011).


\bibitem{Buyukdagli_2013}
{S. Buyukdagli and T. Ala-Nissila} Phys. Rev. E  $\bold{87}$, 063201 (2013).

\bibitem{Andelman_1997}
{Borukhov I., Andelman D., Orland H.} PRL $\bold{79}$, 435 (1997).

\bibitem{Antypov_2005}
{Dmytro Antypov, Marcia C. Barbosa, Christian Holm} PRE $\bold{71}$, 061106, (2005).

\bibitem{Kornyshev}
{Kornyshev A.} J. Phys. Chem. B $\bold{111}$, 5545 (2007).

\bibitem{Maggs2015}
{A. C. Maggs and R. Podgornik} Soft Matter $\bold{12}$, 1219  (2015).

\bibitem{Buyukdagli}
{S. Buyukdagli and T. Ala-Nissila} EPL $\bold{98}$, 60003 (2012).

\bibitem{Buyukdagli_2}
{S. Buyukdagli, C.V. Achim and T. Ala-Nissila} J. Stat. Mech. $\bold{P05033}$, 1  (2011).

\bibitem{Ben-Yaakov_2011}
{Dan Ben-Yaakov, David Andelman, and Rudi Podgornik} J. Chem. Phys. $\bold{134}$, 074705 (2011) .

\bibitem{Andelman_2015}
{Yasuya Nakayama and David Andelman} J. Chem. Phys. $\bold{142}$, 044706 (2015).

\bibitem{Hatlo}
{M. M. Hatlo, R. van Roij and L. Lue} EPL $\bold{97}$, 28010 (2012).

\bibitem{Slavchov}
{Slavchov R.I.} J. Chem. Phys. $\bold{140}$, 164510 (2014).

\bibitem{Budkov}
{Budkov Yu.A., Kolesnikov A.L., Kiselev M.G.} EPL $\bold{111}$, 28002 (2015).

\bibitem{Barrat_Hansen}
{Barrat J.-L., Hansen J.-P.} {\sl Basic concepts for simple and complex liquids} (University Press, Cambridge, 2003).

\bibitem{Sanchez1976}
{Isaac C. Sanchez and Robert H. Lacombe} The Journal of Physical Chemistry $\bold{80}$ (2), 1 (1976).

\bibitem{Fedorov2008}
{Maxim V. Fedorov, Alexei A. Kornyshev} Electrochimica Acta $\bold{53}$, 6835 (2008).

\bibitem{Howard2010}
{Jesse J. Howard, John S. Perkyns, and B. Montgomery Pettitt} J. Phys. Chem. B $\bold{114}$, 6074 (2010).

\bibitem{Schroder_2015}
{Christian Schroder and Othmar Steinhause} J. Chem. Phys. $\bold{142}$, 064503 (2015).

\bibitem{Cavalcante_2014}
{Ary de Oliveira Cavalcant, Mauro C. C. Ribeir and Munir S. Skaf} J. Chem. Phys. $\bold{140}$, 144108 (2014).

\bibitem{Booth1951}
{F. Booth} J. Chem. Phys. $\bold{142}$, 391 (1951).

\bibitem{Sutman1998}
{G. Sutmann} J. Electroanal. Chem. $\bold{450}$, 289 (1998).

\bibitem{Yeh1999}
{In-Chul Yeh and Max L. Berkowitz}  J. Chem. Phys. $\bold{110}$, 7935 (1999).

\bibitem{Gongadze2015}
{Ekaterina Gongadze, Ales Iglic} Electrochimica Acta $\bold{178}$, 541 (2015).

\bibitem{Luczak2013}
{Justyna Luczak, Christian Jungnickel, Marta Markiewicz, and Jan Hupka}  J. Phys. Chem. B, $\bold{117}$, 5653 (2013).
\end{thebibliography}
\end{document}